\documentclass[11pt]{wlscirep}
\usepackage{graphicx}
\usepackage{hyperref}
\usepackage[T1]{fontenc}

\title{Sudden quench of harmonically trapped mass-imbalanced fermions}

\author[1,*]{Dillip K. Nandy}
\author[2]{Tomasz Sowi\'{n}ski}
\affil[1]{Center for Theoretical Physics of Complex Systems, Institute for Basic Science (IBS) \\ Daejeon 34126, Korea}
\affil[2]{Institute of Physics, Polish Academy of Sciences, Aleja Lotnik\'{o}w 32/46, PL-02668, Warsaw, Poland}
\affil[*]{nandy@ibs.re.kr}

\begin{abstract}
Dynamical properties of two-component mass-imbalanced few-fermion systems confined in a one-dimensional harmonic trap following a sudden quench of interactions are studied. It is assumed that initially the system is prepared in the non-interacting ground state and then, after a sudden quench of interactions, the unitary evolution is governed by interacting many-body Hamiltonian. By careful analysis of the evolution of the Loschmidt echo, density distributions of the components, and entanglement entropy between them, the role of mass imbalance and particle number imbalance on the system's evolution stability are investigated. All the quantities studied manifest a dramatic dependence on the number of heavy and lighter fermions in each component at a given quench strength. The results may have implications for upcoming experiments on fermionic mixtures with a well-defined and small number of particles.
\end{abstract} 

\begin{document}  
\flushbottom
\maketitle

\section{Introduction}
Detailed studies of non-equilibrium dynamics of strongly correlated quantum systems have drawn tremendous attention over the last decades due to successful experimental realizations of different artificial many-body quantum models with appropriately tailored ultracold quantum gases~\cite{Bloch2008, Trotzky2012, Georgescu2014,Polkovnikov2011}. With these contemporary experimental setups it is possible to prepare microscopic systems with several degrees of tunability, and thus, simulate different static and dynamical properties of quantum systems in a controlled manner. Importantly, these state-of-the-art experiments make it possible to change external parameters on time scales that could be much faster or slower than the intrinsic microscopic time scales of the system. This possibility opens a route for fundamental studies of a dynamical response to a whole spectrum of external disturbances~\cite{Pustilnik2006,Nascimbene2009, Pereira2009, Polkovnikov2011,2014SindonaNJP,2018BhattacharyaJPhysB,2022DasPhysicaB}.
 
Experiments involving ultracold atoms allow to realize of two-component many-body quantum systems with flavors obeying different quantum statistics, {\it e.g.}, Bose-Bose \cite{SchulzePRA2018,Catani, Thalhammer}, Bose-Fermi \cite{2016OnofrioReview,2002HadzibabicPRL,2006GunterPRL,2009BestPRL,2011WuPRARapid,2014TungPRL,2018LousPRL}, and Fermi-Fermi \cite{Wille,CetinaPRL2015,RavensbergenPRA2018,tiecke2010Feshbach6Li40K,cetina2016ultrafast} mixtures. It is worth noting that also multi-component systems are currently under experimental control \cite{Taglieber1,Taglieber2}. Additionally, these high-precision experiments also offer great tunability of the number of particles in each component and the inter-component interaction strength between the species~\cite{Serwane, ZurnPRL2012,Chin}. This reopened new windows for theoretical investigations of such systems (for review see~\cite{2012BlumeRPP,2016ZinnerEPJ,2019SowinskiRPP,2022MistakidisReviewARX,
Giri2022}). 

Having all these experimental and theoretical scenarios in mind, here we focus on quench dynamics of two-component mass-imbalanced fermionic systems in the regime of a small number of particles. In this way, we try to bridge studies on large ultra-cold mass-imbalanced fermionic mixtures with recent experiments on systems containing several fermions of the same mass in quasi-one-dimensional confinement~\cite{ZurnPRL2012}. Specifically, we investigate the sudden quench dynamics of such systems prepared initially in the noninteracting many-body ground state. We assume that interactions are suddenly switched on and we study the nonequilibrium effect of this sudden perturbation on the system's dynamical properties. As a figure of merit, we use the Loschmidt echo \cite{Campbell, Garcia2} as an adequate indicator for quantifying such a nonequilibrium effect. The Loschmidt echo has been used as an important and very useful tool in many physical out-of-equilibrium dynamical problems. This includes, studies of criticality in spin chains \cite{Quan, Zanardi, Benini2021}, dynamical quantum phase transition in one-dimensional models \cite{Lacki, Heyl}, finite temperature quantum phase transitions \cite{Mera}, orthogonality catastrophe in Fermi polaron system \cite{Knap}, Bose polaron in small quenched quantum gas system \cite{Campbell}, and in many others. Further, we support the conclusions obtained from the Loschmidt echo by examining the evolution of the particle density distributions and the inter-component correlations. Particularly, the effect of the mass imbalance and the particle imbalance are systematically investigated for different systems with up to six particles. To obtain numerically credible results, to find the time-evolved state of the system, we use a combination of methods based on the exact diagonalization of the many-body Hamiltonian and the decomposition of the unitary evolution operator into the Chebyshev polynomials (so-called, the Chebyshev evolution technique) \cite{HalimehPRB2019, Leforestier1991}. In this way, we can obtain well-converged results for short-time as well as long-time dynamics. 

Here we want to emphasize that interaction-driven quench processes have also been studied previously for two-component mixtures of different statistics, particularly systems consisting of few particles~\cite{Campbell,2015MistakidisPRA,2017MistakidisPRA,2018MistakidisNJP,Erdmann2019}. For instance, the authors considered a few-body system consisting of two bosons and a single impurity atom of equal mass. They started with an initial state in which the two bosons are in the Tonks-Girardeau interacting limit and then study the evolution of the Loschmidt echo at different inter-component interaction strengths. This study merely implies that even a small mixture of quantum particles can witness complete orthogonality between two states evolved by the Hamiltonians before and after a sudden quench. Similarly, various dynamical properties for interaction quench-driven processes in systems with two bosonic impurities immersed in a Fermi sea for a mass-balanced scenario were recently studied~\cite{Mukherjee}.  
 
\section{The system studied}
In our present model, we focus on a few-body system containing two-component mass-imbalanced spin-polarized fermions confined in a one-dimensional harmonic trap. Particles from opposite components interact with each other by short-range contact forces through a low-energy ($s$-wave) scattering channel. The many-body Hamiltonian representing this mesoscopic system can then be given by
\begin{eqnarray} \label{Hamiltonian}
 \hat{\mathcal{H}}(g) = \sum_{\sigma} \int\!\!\mathrm{d}x\,\hat{\Psi}^{\dagger}_{\sigma}(x) H_{\sigma} \hat{\Psi}_{\sigma}(x)
+ g \int \mathrm{d}x\,\hat{\Psi}^{\dagger}_{\uparrow}(x) \hat{\Psi}^{\dagger}_{\downarrow}(x) \hat{\Psi}_{\downarrow}(x)  
 \hat{\Psi}_{\uparrow}(x),
\end{eqnarray}
where $\hat{\Psi}_{\sigma}(x)$ denotes the field operator annihilating fermion from a component $\sigma\in\{\downarrow,\uparrow\}$ at a spatial position $x$, while $g$ is the effective short-range interaction strength between components~\cite{OlshaniiPRL1998}. $H_\sigma$ represents the single-particle Hamiltonian for the component $\sigma$ and it is justified in the following. It is straightforward to check that the Hamiltonian (\ref{Hamiltonian}) does commute with the particle number operators for each component, $\left[\hat{\cal H}(g), \hat{N}_\sigma \right] = 0$, where $\hat{N}_\sigma=\int\mathrm{d}x\hat{\Psi}_{\sigma}^\dagger(x)\hat{\Psi}_{\sigma}(x)$. Thus, its properties can be explored in the subspaces of fixed numbers of particles $N_\uparrow$ and $N_\downarrow$. For convenience, in our work, we use the notation $[\![N_\uparrow,N_\downarrow]\!]$ to specify the number of particles in each component, by $N=N_\uparrow+N_\downarrow$ we indicate the total number of particles, and we write $\hat{\cal H}_0$ for the non-interacting Hamiltonian $\left.\hat{\cal H}(g)\right|_{g=0}$.

We focus on mixtures of fermions having different masses which currently are under investigation in various experimental laboratories. In particular, we study quench dynamics with our few-body setup for mass ratios $\mu\in\{1.8, 2.2, 4.0, 4.2, 6.7\}$ which correspond respectively to the following ultracold atomic mixtures: $^{161}$Dy-$^{87}$Sr, $^{87}$Sr-$^{40}$K, $^{161}$Dy-$^{40}$K, $^{167}$Er-$^{40}$K, $^{40}$K-$^{6}$Li. In our theoretical model, we assume that both fermionic components experience the same one-dimensional trapping potential. This commensurate potential approximation can also be realized in experiments by properly tuning the width and intensity of the applied laser field. Of course, analogous calculations can be carried out for systems with component-dependent potentials. However, in the present study, we mostly focus on the effect of mass and particle imbalance on the dynamical properties of the system, so our approximated model can capture most of the relevant physics.  

With the above approximation, the single-particle Hamiltonian takes the following form 
\begin{equation}
H_\sigma = -\frac{\hslash^2}{2m_\sigma}\frac{\mathrm{d}^2}{\mathrm{d}x^2}+\frac{m_\sigma \omega^2}{2}x^2.
\end{equation}
By expressing all the physical quantities in the natural units of the harmonic oscillator related to one of the flavors (here we choose a lighter spin-down component) the single-particle Hamiltonians can be simplified as
\begin{eqnarray} 
 H_{\downarrow} &= -\frac{1}{2}\frac{\mathrm{d}^2}{\mathrm{d}x^2} +\frac{1}{2}x^2 , \label{HamDown} ~~~~
 H_{\uparrow} &= -\frac{1}{2\mu}\frac{\mathrm{d}^2}{\mathrm{d}x^2} +\frac{\mu}{2}x^2, \label{HamUp}
\end{eqnarray}
where $\mu=m_\uparrow/m_\downarrow$ is the mass ratio between two fermionic flavors. Obviously, analytical expressions for all eigenstates of these differential operators are well-known and given by 
\begin{subequations}
\begin{eqnarray}
 \phi^{\downarrow}_n (x) &= \left( 2^n n! \sqrt{\pi}\right)^{-1/2} \mathrm{H}_{n}(x) e^{-x^2/2}, \\ 
 \phi^{\uparrow}_n (x) &= \left( 2^n n! \sqrt{\pi/\mu}\right)^{-1/2} \mathrm{H}_n(\sqrt{\mu}x) e^{-\mu x^2/2},
\end{eqnarray}
\end{subequations}
where $\mathrm{H}_n (x)$ is the $n$-th order Hermite polynomial. In terms of these single-particle orbitals one can now introduce corresponding annihilation operators $\hat{a}_n$ and $\hat{b}_n$ (respectively for lighter and heavier component particles) to expand the field operators as
\begin{eqnarray} \label{Decomposition}
 \hat{\Psi}_{\downarrow}(x) = \sum_n \hat{a}_{n} \phi^{\downarrow}_{n}(x), \qquad
 \hat{\Psi}_{\uparrow}(x) = \sum_n \hat{b}_{n} \phi^{\uparrow}_{n}(x),
 \end{eqnarray}
and write the original Hamiltonian (\ref{Hamiltonian}) as
\begin{eqnarray}
\hat{\mathcal{H}}(g) = \sum_i \varepsilon_{i}\left( \hat{a}^{\dagger}_{i}\hat{a}_{i} + \hat{b}^{\dagger}_{i}\hat{b}_{i}\right) + g\sum_{ijkl} U_{ijkl} \hat{a}^{\dagger}_{i}\hat{b}^{\dagger}_{j}\hat{b}_{k}\hat{a}_{l}
\label{Ham} 
\end{eqnarray}
Note that in the case studied the single-particle energy $\epsilon_i=i+1/2$ is the same for both components. The two-body interaction coefficients are given by the general formula  
\begin{eqnarray}
 U_{ijkl} = \int \mathrm{d}x \ \phi^{\uparrow *}_{i}(x) \ \phi^{\downarrow *}_{j}(x) \ \phi^{\downarrow}_{k}(x) \ \phi^{\uparrow}_{l}(x).
\end{eqnarray}
and in the case of harmonic oscillator confinement, they can be calculated analytically~\cite{2020RojoMathematics}. In this way one may represent the original Hamiltonian (\ref{Hamiltonian}) as a matrix with matrix elements of the generic form ${\cal H}_{ij}=\langle {\cal F}_i|\hat{\cal H}(g)|{\cal F}_j\rangle$, where the set $\{|{\cal F}_i\rangle\}$ contains all possible Fock states representing non-interacting Hamiltonian ${\cal H}_0$. In principle, this approach allows one to perform all calculations in this representation. In practice, however, some reasonable cutoff of the Fock basis is needed due to a finite numerical resources used~\cite{2020RojoMathematics,1998HaugsetPRA,2017RaventosJPhysB,2019ChrostowskiAPPA}.

\begin{figure}[t!] 
\centering
\includegraphics[width=0.8\linewidth]{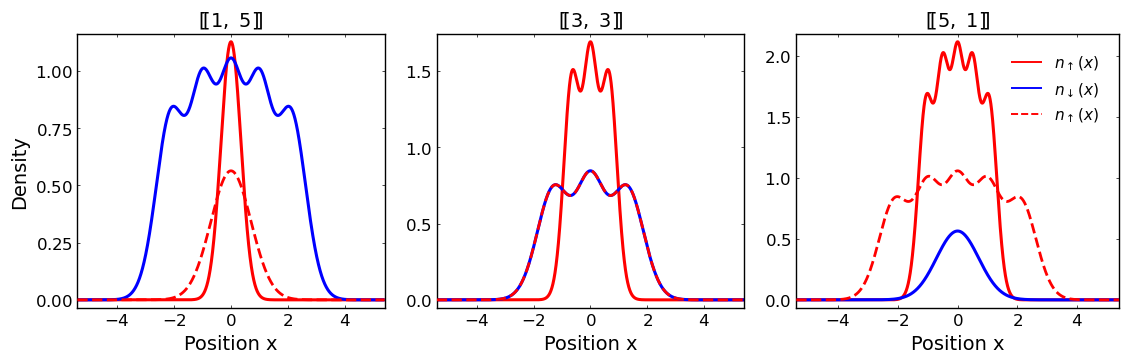}
\caption{Initial single-particle density distribution $n_{\sigma}(x) = \langle \mathtt{ini}|\hat{\Psi}_{\sigma}^{\dagger}(x) \hat{\Psi}_{\sigma}(x) |\mathtt{ini}\rangle$ of components for $N=6$ particles and different imbalances with fixed mass ratio $\mu=4$. For comparison, with the red dashed line, we indicate the distribution of the heavy component in the limit of equal masses, $\mu \rightarrow 1$. Note, that increasing mass ratio changes the overlap of density distributions and in consequence, it affects the interaction energy. This change is more pronounced for a larger number of heavy particles present in the system.} 
\label{Fig1}
\end{figure}
In our work, we focus on the dynamical properties of the system after a sudden quench of interactions. We assume that initially the system is prepared in the non-interacting ground state of the system $|\mathtt{ini}\rangle$ with eigenenergy $E_\mathtt{ini}=(N_\uparrow^2+N_\downarrow^2)/2$.
For convenience of further discussion in Fig.~\ref{Fig1} we present initial density distribution $n_{\sigma}(x) = \langle \mathtt{ini}|\hat{\Psi}_{\sigma}^{\dagger}(x) \hat{\Psi}_{\sigma}(x) |\mathtt{ini}\rangle$ of different components for exemplary case of $N=6$ particles differently distributed between components. Then, at $t=0$, the interaction strength is suddenly changed to $g$ and the system starts to evolve according to the Hamiltonian $\hat{\cal H}(g)$. In general, the most straightforward method of finding the time-evolved state for short times is to solve time-dependent Schr\"odinger equation, $i\partial_t|\Psi(t)\rangle= \hat{\cal H}(g)|\Psi(t)\rangle$, with the standard fourth-order Runge-Kutta algorithm. If the long-time behavior of the system is considered, then one first needs to perform exact diagonalization of the target Hamiltonian $\hat{\cal H}(g)$. Then the time-evolved state can be obtained as an appropriate superposition of time-evolved eigenstates $\{|\Upsilon_n \rangle \}$ having corresponding eigenenergies $E_n$
\begin{equation} \label{standardevolution}
|\Psi(t)\rangle = \sum_n \langle\Upsilon_n|\mathtt{ini}\rangle\,\mathrm{e}^{-iE_nt}\,|\Upsilon_n\rangle
\end{equation}
In practice, these two straightforward methods are however feasible only for relatively small systems\cite{2018KehrbergerPRA} or for large systems close to the non-interacting limit. It comes from the fact that the size of the Hilbert subspace of states which have a relevant contribution to the dynamics strongly depends on interaction strength $g$ and also grows exponentially with the number of particles $[\![N_\uparrow,N_\downarrow]\!]$. Therefore, to tackle larger systems we employ the Chebyshev time evolution technique, which has been successfully applied to many physical problems concerning time dynamics \cite{Garcia3, Halimeh}. We found that the Chebyshev evolution technique is a quite remarkable approach to producing short-time and long-time dynamical behavior. The Chebyshev time-evolution technique is based essentially on expressing the unitary evolution operator $U(t) = \mathrm{e}^{-i \hat{\cal H}(g) t}$ in terms of Chebyshev polynomials $\mathrm{T}_k(x)$. The technical details of the method and relevant derivations can be found in \cite{Halimeh, Dobrovitski, Leforestier}. In the following, we present only working steps in this framework. The formal form of the expansion reads
\begin{equation} \label{ChebDecomp}
 U(t) = e^{-i\beta t} \left( c_0 + 2\sum^{\mathcal{D}}_{k=1} c_k(t)\,\mathrm{T}_k ({\hat{\cal R}}) \right),
\end{equation}
where $\hat{\cal R}=\left[\hat{\cal H}(g) - \beta\right]/\alpha$ is the so-called rescaled Hamiltonian. Simply, this is the Hamiltonian shifted and rescaled in such a way that the relevant part of its spectrum is spanned between $[-1, 1]$. The scaling parameters $\alpha$ and $\beta$ are determined by extreme eigenvalues ($E_\mathrm{min}$ and $E_\mathrm{max}$, respectively) of the original Hamiltonian $\hat{\cal H}$ and they are defined as $\alpha = (E_\mathrm{max}-E_\mathrm{min})/2$ and $\beta = (E_\mathrm{max} + E_\mathrm{min})/2$.

In contrast to the standard method for time evolution \eqref{standardevolution}, application of the Chebyshev time-evolution technique does not require knowledge of all eigenstates and eigenvalues of the many-body Hamiltonian, but only extremal eigenenergies $E_\mathrm{min}$ and $E_\mathrm{max}$. Since they can be found quite quickly with a few iterations of the standard Lanczos algorithm, the method becomes feasible also for relatively large Hamiltonian matrices. Moreover, the accuracy of this approach is controlled solely by the order of the Chebyshev polynomial $\cal D$. It is set at a reasonably large value to get a well-converged dynamical behavior. At this point, one should remember that the full many-body Hamiltonian \eqref{Hamiltonian} is in principle not bounded from above. Therefore, maximal energy $E_\mathrm{max}$ is not well-defined. However in practice, due to a specific cut-off of the Fock space used to represent the Hamiltonian as a finite matrix, the largest eigenvalue $E_\mathrm{max}$ exists and it is directly related to this cut-off. In our calculations, the cut-off is determined by the number of single-particle orbitals used in the decomposition \eqref{Decomposition}. For systems with a single impurity and two, three, four, and five particles in the majority component we take respectively 20, 20, 18, and 15 single-particle orbitals. For the balanced systems with two, four, and six particles we take 25, 18, and 12 orbitals, respectively. Note also that extremal energies $E_\mathrm{min}$ and $E_\mathrm{max}$ is not unique for all systems since they depend also on the mass ratio $\mu$. We checked that further increasing the number of orbitals affects the dynamics only minutely.

The expansion coefficients $c_k(t)$ in the decomposition (\ref{ChebDecomp}) are given by $c_k = (-i)^k \mathrm{J}_k(\alpha t)$, where $\mathrm{J}_k(\tau)$ denotes the $k$-th order spherical Bessel function of the first kind. The maximum order of the Chebyshev polynomial is determined by the dimension of the cropped Hilbert space $\mathcal{D}$. In practice, however, due to a superexponential decay of Bessel functions $\mathrm{J}_k(\tau)\sim (\tau/k)^k$, one can truncate the summation on a much smaller cutoff ${\cal D}_\mathrm{C}$ after achieving the desired accuracy. For example, for $\mathcal{D} = 48400$ (corresponding to $N_\uparrow=N_\downarrow=3$ with ${\cal K}=12$ single-particle orbitals for each component), we find that ${\cal D}_C \sim 400$ is sufficient to ensure numerical convergence of the results for the whole time domain considered in our work.

\section{Loschmidt echo}
One of the most relevant figures of merit for capturing the nonequilibrium effect of a quenching process is the dynamical Loschmidt echo $\mathcal{L}(t)$ commonly defined as \cite{Campbell, Garcia2}
\begin{equation}
 \mathcal{L}(t) = |\langle \mathtt{ini}| \mathrm{e}^{i \hat{\cal H}_0 t}|\Psi(t) \rangle|^2 = |\langle \mathtt{ini}| \mathrm{e}^{i \hat{\cal H}_0 t} \mathrm{e}^{-i \hat{\cal H}(g) t}|\mathtt{ini} \rangle|^2.
\end{equation}
\begin{figure}[t!]
\centering
\includegraphics[width=0.8\linewidth]{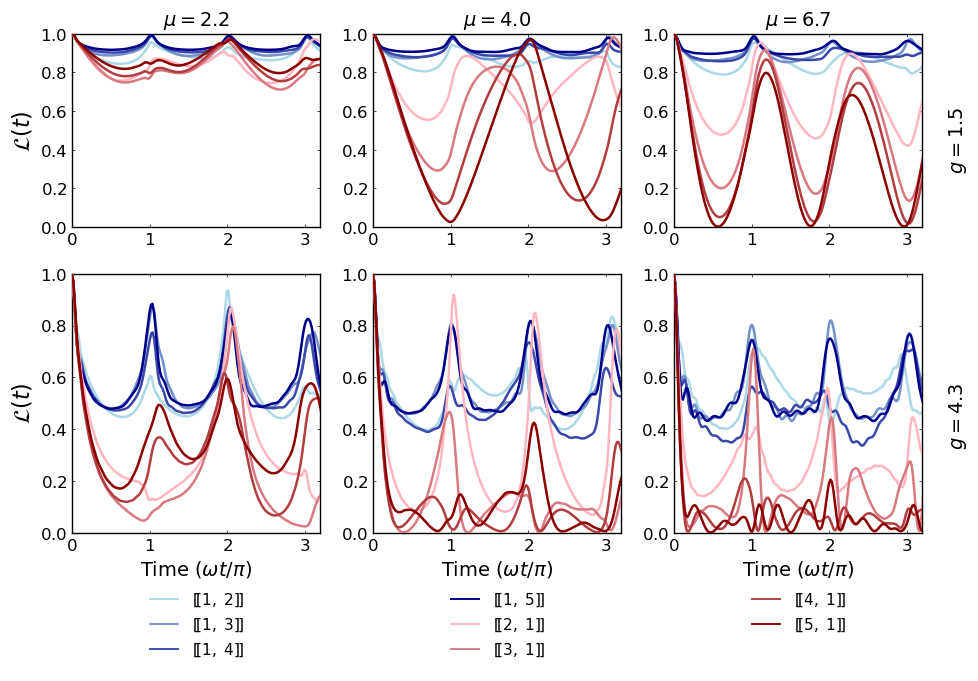}
\caption{Time evolution of the Loschmidt echo $\mathcal{L}(t)$ of single-impurity systems containing a different number of particles (blue and red lines respectively for systems $[\![1,N_\downarrow]\!]$ and $[\![N_\uparrow,1]\!]$) calculated for two different interactions (rows) and three different mass ratio (columns). Notice that systems with lighter impurity are much more sensitive than their heavy-impurity counterparts.} 
\label{Fig2}
\end{figure}
From the above formulation, it is clear that $\mathcal{L}(t)$ essentially quantifies differences between the initial state $|\mathtt{ini}\rangle$ unitary time-evolved by the Hamiltonian $\hat{\cal H}(g)$ and by the non-interacting Hamiltonian $\hat{\cal H}_0$. Thus, whenever the echo becomes close to one, the time-evolved state of an interacting system is well-approximated by the state obtained with a trivial evolution of the non-interacting system with non-interacting Hamiltonian $\hat{\cal H}_0$. Using eigenstates $\{|\Upsilon_n \rangle \}$ of the final Hamiltonian $\hat{\cal H}(g)$ and corresponding eigenenergies $E_n$, one can further simplify $\mathcal{L}(t)$ as 
\begin{equation}
 \mathcal{L}(t) = \left|\sum_n \mathrm{e}^{i(E_n - E_\mathtt{ini})t} |\langle \mathtt{ini} |\Upsilon_n \rangle|^2 \right|^2.  
\label{LE}         
\end{equation}
This reformulation gives another interpretation of the Loschmidt echo as a cumulative interference of interacting eigenstates contributing to the initial state. However, calculation of $\mathcal{L}(t)$ with this definition is in practice not feasible for larger systems since it requires precise knowledge of the full eigenspectrum of interacting Hamiltonian $\hat{\cal H}(g)$. As such, it is suitable rather for smaller system sizes with a reasonable number of single-particle orbitals.  

Let us first discuss the dynamical behavior of the Loschmidt echo for different mass mixtures containing an imbalanced number of particles. In the simplest case, we consider a single impurity (lighter or heavier) immersed in a few-fermion bath of the opposite component. In Fig.~\ref{Fig2}, we present the time dependence of ${\cal L}(t)$ for two different interaction strengths $g$ and three different mass ratios $\mu$. Different blue and red shaded lines correspond to systems with heavier and lighter impurity and a different number of majority particles ({\it i.e.}, systems with $[\![1,N_\downarrow]\!]$ and $[\![N_\uparrow,1]\!]$ particles, respectively). As a general observation, we notice that whenever interactions between the lighter impurity with a heavy fermionic bath are switched on the Loschmidt echo decays quite faster than in the opposite situation with a single heavy impurity immersed in a lighter medium. It is also much more sensitive to the value of mass ratio and even for relatively weak interactions it may rapidly drop to zero. This suggests that small polaron systems with light impurity are very sensitive to interactions and evolve significantly different than their non-interacting counterparts. In contrast, systems with heavy impurity typically are very robust to interactions and also to the mass difference. It is clear, that independently of the mass ratio, all blue lines behave very similarly. Thus, an increasing number of particles in the majority component has the opposite impact on the Loschmidt echo. In contrast, for lighter impurity (system with $[\![N_\uparrow,1]\!]$ particles), when the number of particles in the bath is increased, the dynamics of the system drives away more deeply from the non-interacting case. The main reason standing behind such different resistivity of light and heavy impurity comes from the essential difference of initial density distribution profiles (see Fig.~\ref{Fig1}). In the case of heavy impurity, the increasing number of lighter particles almost does not affect the interaction energy (proportional to the overlap of single-particle density profiles). Thus it changes the behavior of the system only marginally. The situation is significantly different when light impurity is considered. Then, the density overlap changes significantly when the number of heavy particles is increased. Provided that the mass ratio is sufficiently large, the density distribution of the light particle is always spread over the entire heavy bath.
\begin{figure}[t!]
\centering
\includegraphics[width=0.8\linewidth]{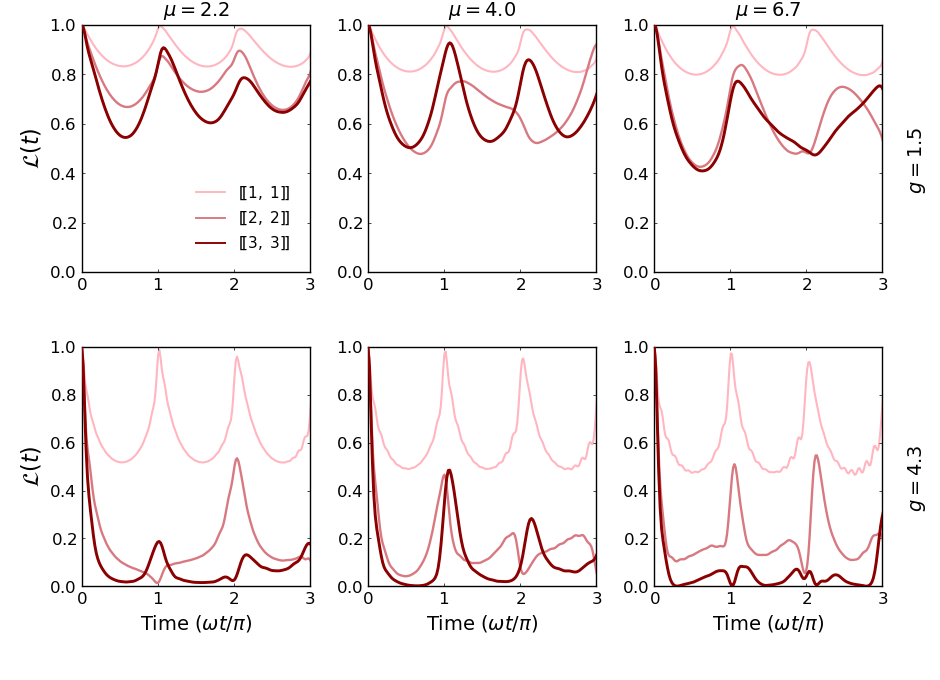}
\caption{Time evolution of the Loschmidt echo $\mathcal{L}(t)$ of systems with an equal number of particles in both components, $[\![N/2,~ N/2]\!]$. Clearly, the behavior of the minimal system with $N=2$ particles $[\![1,~ 1]\!]$ is independent of the mass imbalance at a given quench strength. For larger systems, the Loschmidt echo drops faster and deeper for systems with stronger interactions.} 
\label{Fig3}
\end{figure}

\begin{figure}[t!]
\centering
\includegraphics[width=0.8\linewidth]{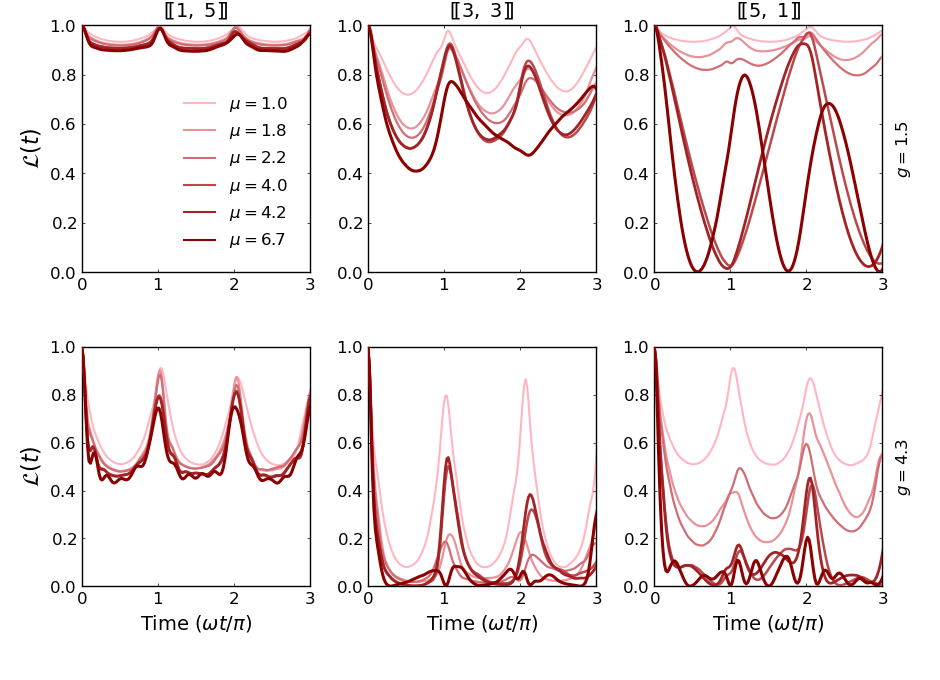}
\caption{Time evolution of the Loschmidt echo $\mathcal{L}(t)$ calculated for systems containing $N=6$ particles and different imbalances (particle and mass). Clearly, the heavy-impurity system (left column) is highly resistant to the sudden perturbation irrespectively of the mass imbalance. Along with the increasing number of heavy particles, $\mathcal{L}(t)$ gradually becomes more affected by mass-imbalance.} 
\label{Fig4}
\end{figure}

An intermediate behavior of $\mathcal{L}(t)$ is clearly visible for systems with a balanced number of particles, $[\![N/2,N/2]\!]$ (see Fig.~\ref{Fig3}). Along with the increasing number of particles, interactions, and/or mass ratio, the system drives away from the noninteracting dynamics and its periodic revivals become less accurate. Only a fully integrable case with $[\![1,1]\!]$ particles manifests relative stability and resistance to the model's parameter change. For completeness of this analysis, in Fig.~\ref{Fig4} we expose an impact of the mass difference on the dynamical properties of the system with $N=6$ particles differently distributed among components. The detailed inspection and comparison show a general trend that increasing of mass ratio leads to a stronger and quicker departure from the non-interacting dynamics. This effect is further enhanced when the number of heavier atoms increases. Indeed, the time evolution of the Loschmidt echo for the system with only one heavy atom (left column in Fig.~\ref{Fig4}) is poorly sensitive to the mass ratio $\mu$. Contrary, in the case of a system with a single lighter impurity (right column in Fig.~\ref{Fig4}) the echo quickly drops close to zero for a large enough mass ratio even for small interaction strengths. 
\begin{figure}[t!]
\centering
\includegraphics[width=0.8\linewidth]{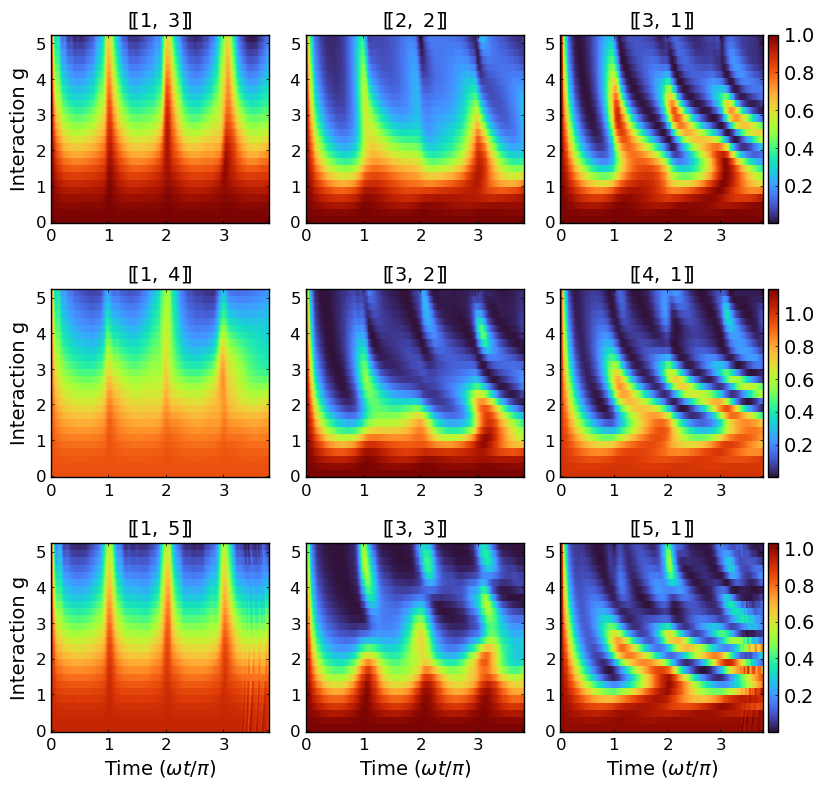}
\caption{Dependence of time evolution of the Loschmidt echo ${\cal L}(t)$ on interactions calculated for systems with a different number of particles and fixed mass ratios $\mu=4.0$. It is clearly visible that the dynamical response of the system is regular for heavy-impurity systems (left column) and become strongly unpredictable (for sufficiently large interactions) when particle imbalance is shifted towards systems with a smaller number of light particles (right column). } 
\label{Fig5}
\end{figure}
Further, to understand the role of the number of particles and particle imbalance with respect quench strength, in Fig.~\ref{Fig5} we present the full dynamical picture of the Loschmidt echo for the whole range of interactions after the quench for systems with $N=4$, $5$, and $6$ particles and fixed mass ratio $\mu=4.0$ (dysprosium-potassium mixture). It is clear that the most stable and regular pattern is obtained for systems with only one heavy impurity $[\![1,N_\downarrow]\!]$. In these cases revivals (${\cal L}\approx 1$) appear periodically for any interaction strength and every number of lighter particles. Situation changes, when the number of heavy particles increases. For example, for balanced systems $[\![N/2,N/2]\!]$ regular pattern is present only for smaller interactions and a smaller number of particles. In the case of an extreme case, when only one light particle is present in the system ({\it i.e.} the system contains $[\![N_\uparrow,1]\!]$ particles), the Loschmidt echo pattern quickly become irregular, and in fact unpredictable. 

\begin{figure}[t!]
\centering
\includegraphics[width=0.8\linewidth]{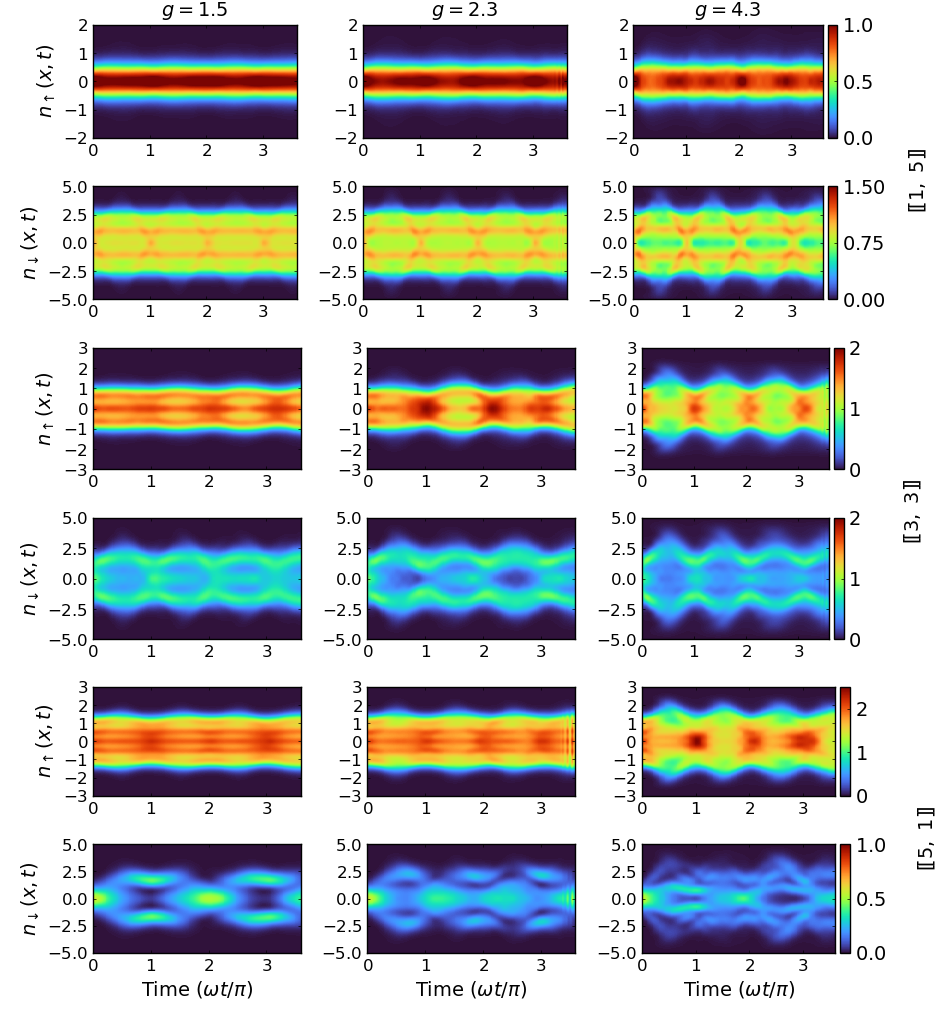}
\caption{Time evolution of the single-particle density distributions $n_\sigma(x,t)$ for six-particle systems with different imbalances and interactions. Here, to keep correspondence with Fig.~\ref{Fig5}, mass ratio is fixed $\mu=4.0$.} 
\label{Fig6}
\end{figure}

\section{Density profile after quench}
The dynamical behavior of the system after the quench of interactions is reflected in many different, directly measurable quantities. One of them is the single-particle density profile which is defined as 
\begin{equation}
 n_{\sigma} (x,t) = \langle \Psi (t)|\hat{\Psi}_{\sigma}^{\dagger}(x) \hat{\Psi}_{\sigma}(x) |\Psi (t) \rangle.
\end{equation}
The time evolution of these densities for both components is presented in Fig.~\ref{Fig6} for differently balanced systems containing $N=6$ particles. For clarity, we present calculations only for one selected mass ratio $\mu = 4.0$ (dysprosium-potassium mixture) and three different interaction strengths. Nevertheless, we checked that calculations made for other parameters lead to analogous conclusions. 

For an imbalanced system with one heavy impurity $[\![1,5]\!]$, we notice that the spatial extent of the impurity is mildly affected by the interactions and the profile stays highly Gaussian around the harmonic center. Only for strong enough interactions, some sort of fragmentation of the density occurs and some deviations from the Gaussian shape start to build up. Subsequently, the effect of the heavy particle on the lighter medium is also not significant, leading to an almost stationary density distribution of the majority component overall range of interactions studied. Such a mild effect of interaction quench on the density profile of both components is fully compatible with slow and regular changes of the Loschmidt echo $\mathcal{L}(t)$ (compare with Fig.~\ref{Fig2}). On the contrary, in the case of a system containing a majority of heavy particles and a single light impurity $[\![5,1]\!]$, the effect of the sudden quench process is substantial. Time evolution of the density profile of the heavy component $n_\uparrow(x,t)$ is strongly affected and quickly become unpredictable. The density distribution of light impurity $n_\downarrow(x,t)$ is affected even much stronger and smeared over the entire area in which the system is located. Exactly as was observed in the case of Loschmidt echo, the behavior of a fully balanced system with $[\![3,3]\!]$ particles is intermediate when compared to the two extreme cases discussed. Again, these different behaviors of the system for a different distribution of particles between components can be viewed as a consequence of substantially different density distributions at the initial moment. It is worth adding, that similar evolution of density pattern was recently found in the case of a spinor impurity atom immersed in a BEC of spinless bosons \cite{Mistakidis2020Bosepolaron}. 

By comparing time evolutions of the Loschmidt echo $\mathcal{L}(t)$ and the density distributions $n_\sigma(x,t)$ one can observe almost ideal coincidence correspondence of the revival moments for these quantities. It is also evident that whenever the impurity atom comes close to the center of the trap, the value of the echo $\mathcal{L}(t)$ increases. Contrary, it is diminished when the impurity moves away from the center. Moreover, the density profiles for the balanced system show a symmetric behavior around the trap center at any value of interaction quench. It is interesting to note that a very similar behavior of the density profile was recently observed for a system containing two bosons and one impurity atom \cite{Campbell}. 

\begin{figure}[t!]
\centering
\includegraphics[width=0.8\linewidth]{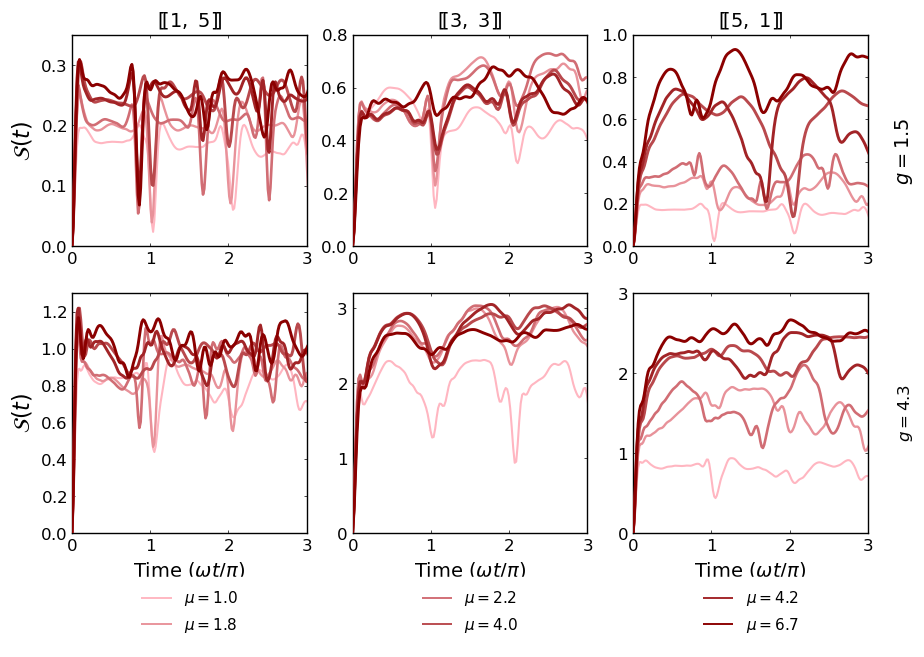}
\caption{Time evolution of the von Neumann entropy $S(t)$ quantifying inter-component correlations in systems of six particles obtained for different mass ratios and different interactions. As one can see, the correlation produced in $[\![1,~ N_{\downarrow}]\!]$ system is much smaller than the other two six-particle systems at the same quench strength. Further, the system with a single impurity manifests stronger dependence on the mass imbalance when compared to the balanced case.} 
\label{Fig7}
\end{figure}

\begin{figure}[t!]
\centering
\includegraphics[width=0.8\linewidth]{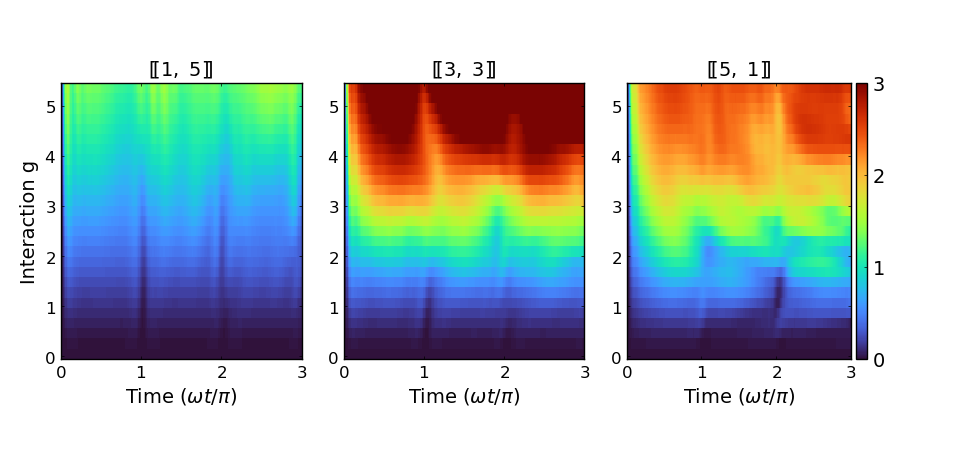}
\caption{Time evolution of the von Neumann entropy $S(t)$ calculated for systems with $N=6$ particles and different imbalances and fixed mass ratios $\mu=4.0$ as a function of quench strength. It is clear that the entanglement production is rather weak and regular for the heavy-impurity system (left column) while becomes enhanced when particle imbalance is shifted towards the light-impurity case (right column). }
\label{Fig8}
\end{figure}

\section{Formation of inter-component correlations}
A complimentary look at the effect of sudden quench on the dynamical properties of the system can be obtained by calculating the time evolution of inter-component correlations. They can be quantified through the von Neumann entropy given by \cite{Campbell}
\begin{equation} \label{entropy}
S(t) = -\mathrm{Tr}\left[\hat{\rho}_{\sigma}(t)\,\mathrm{ln}\hat{\rho}_{\sigma} (t)\right],
\end{equation}
where $\hat{\rho}_{\sigma}(t)$ is the temporal reduced density matrix of the $\sigma$-component calculated by tracing-out opposite component $\sigma'$ from the operator projecting to the temporal state $|\Psi(t)\rangle$, {\it i.e.}, $\hat{\rho}_{\sigma}(t)=\mathrm{Tr}_{\sigma'} \left[|\Psi(t)\rangle\langle\Psi(t)|\right]$. It is a matter of fact that the value of the entropy (\ref{entropy}) does not depend on the choice of the component $\sigma$ used for calculations. However, from the computational point of view, tracing out a larger component is usually much more efficient. Since both components are distinguishable and initially the system is prepared in the product state $|\mathtt{ini}\rangle$, just after the quench the entropy $S(0) = 0$. Then, due to inter-component interactions, entropy starts to increase signaling the appearance of inter-component correlations. Depending on interactions and the number of particles, the formation of correlations has different intensities. As suspected, due to a difference in initial density distributions, inter-particle correlations are built much faster in systems when the number of heavy particles is larger. In Fig.~\ref{Fig7} we present detailed results obtained for systems with $N=6$ particles and two different values of interactions. In addition, in Fig.~\ref{Fig8} we present complete results for the same systems obtained in a whole range of interactions. In this way, it is possible to compare entropy production with the evolution of the Loschmidt echo (bottom row in Fig.~\ref{Fig5}). When relating the temporal evolution of entropy to the characteristic time of a harmonic oscillator, we see that in each case considered, it increases rapidly at the beginning of the evolution, and then stabilizes at a certain level long before the end of the first period. The evolution shows characteristic periodic peaks. However, they are weaker as the interactions become stronger and the mass ratio larger. This is fully consistent with the Loschmidt echo revivals shown earlier. It is also clear that entanglement entropy increases faster for systems with a larger mass ratio $\mu$ (it is particularly visible when a strongly imbalanced system with light impurity is considered). This is also in agreement with previous results.

Let us now finally compare the evolution of the inter-component correlations with the dynamics of spatial distributions of the components. As argued above, for the system with $[\![1,5]\!]$ particles the spatial extent of the heavy impurity is mildly affected by interactions with the lighter environment. Thus, the two components are probably less correlated and subsequently, the value of entropy $S(t)$ saturates on a relatively small level. In contrast, for the system with $[\![5,1]\!]$ particles the density profile of the lighter impurity is highly influenced by the presence of a heavy bath implying strong correlations between components. In consequence, entanglement entropy approaches higher levels. This comparison proves that mutual influences of the components forced by interactions have a quantum nature and cannot be explained by any classical picture of interacting fluids. 

\section{Conclusions}
We have investigated the dynamical properties of a system consisting of a few mass-imbalanced harmonically trapped fermions after a sudden quench of interactions. Comparing the dynamical behavior of the Loschmidt echo $\mathcal{L}(t)$ we show increasing orthogonality to the non-interacting evolution along with increasing mass imbalance and interactions. As suspected, stronger interactions are directly responsible for the faster creation of correlations between particles. This is reflected in the faster production of entanglement and its saturation on a larger level. Consequently, the temporal state of the system faster departs from the time-evolved state for non-interacting cases (reflected by the quick change of the Loschmidt echo). Our less intuitive finding is that the orthogonality strongly depends on the number of particles in the heavy and light components. In particular, a strong enhancement of the echo's decay is observed by increasing the quench strength for the systems with larger heavy components. In these systems, the long-time dynamics become very chaotic, especially for large mass imbalances. A system with a small number of heavy fermions manifests quite strong stability, independently of the mass ratio between particles from opposite components. Intermediate behavior is observed for systems with a balanced number of particles. We attribute this rather significant difference in the behavior of systems with different imbalances mainly to differences in the density distributions of the individual components. These differences significantly change the overlap integrals, which are related to the interaction energy.

The behavior of the Loschmidt echo is further supported by examining the temporal density distributions and inter-component correlations. We find that for systems with a single heavy impurity the interaction quench (even strong) does not change significantly their spatial distributions. In contrast, for systems with light impurity, the temporal density distributions become highly affected by surrounding heavier medium. As a result, also the spatial distribution of the heavier component is significantly enlarged, especially for stronger quenches. Again, for the balanced-particle scenario, the density profiles become diffusive with the increase of the inter-component coupling. However, this spreading occurs quite smoothly maintaining a periodic trend for both flavors even at the strong quench. 

With these results in hand, one can explore further different directions of such mass-imbalanced systems. Here, we give some of the possible future studies that can be worth looking at. As we found the physical properties of the polaron highly depend on the nature of the impurity atom (heavier/lighter) compared to the medium and quantum statistics. Thus, it would be interesting to explore the dynamical orthogonality of small polaron systems in experiments for bosonic particles as well as for flavors having particles of different mass. As an extension, one can study the Loschmidt echo dynamics of such systems for a more general initial state. For instance, considering scenarios in which the system is initially prepared in some superposition of non-interacting states or even in a thermal mixed state may be very instructive and relevant, also from an experimental perspective. One can also rigorously study the effect of an increasing number of impurity atoms (multi-polaron systems) on orthogonality and consequently capture dynamical effects forced by quantum statistics. It is also worth investigating the quasiparticle excitations properties through radio-frequency spectroscopy or analyzing the spectral function from $\mathcal{L}(t)$ of the present confined system. Another useful direction is to unravel the few to many-body crossover point concerning the size of the bath component both for $[\![1,N_{\downarrow}]\!]$ and $[\![N_{\uparrow},1]\!]$ scenario. Finally, our present study can be a starting point for investigating the dynamical behavior of the impurity subjected to a quantum quench process in a one-dimensional optical lattice focusing on a mass-imbalanced two-component fermionic system with harmonic confinement. Such an optical lattice system can provide a platform to understand the polaron physics in an extended system on a general ground compared to its equal mass counterpart.   

\section*{Acknowledgements}
D.K.N. gratefully acknowledges the funding support by the Institute for Basic Science in Korea (Grant No. IBS-R024-D1). D.K.N. also acknowledges the use of the high-performance computing facility (FERMI cluster) at IBS-PCS.

\section*{Author contributions statement}
D.K.N., and T.S. equally contributed in all stages of the project. All authors reviewed the manuscript.

\section*{Data availability}
All data generated or analyzed during this study are available from the corresponding author upon reasonable request.

\section*{Additional information}

\textbf{Competing financial and non-financial interests} All authors declare no competing interests.

\bibliography{biblio_Lecho}
\end{document}